\documentclass[prl,aps,twocolumn,showpacs]{revtex4}
\usepackage{epsfig,latexsym}
\usepackage{color}
\usepackage[hypertex]{hyperref}

\begin{document}
\title{Contest based on a directed polymer in a random medium}

\author{Cl\'ement Sire}
\email{clement.sire@irsamc.ups-tlse.fr} \affiliation{Laboratoire de
Physique Th\'eorique -- IRSAMC, CNRS\\ Universit\'e de Toulouse, 31062
Toulouse, France}


\begin{abstract}
We introduce a simple one-parameter game derived from a model
describing the properties of a directed polymer in a random medium.
At his turn, each of the two players picks a move among two
alternatives in order to maximize his final score, and minimize
opponent's return. For a game of length $n$, we find that the
probability distribution of the final score $S_n$ develops a
traveling wave form, ${\rm Prob}(S_n=m)=f(m-v\,n)$, with the wave
profile $f(z)$ unusually decaying as a double exponential for large positive
and negative $z$. In addition, as the only parameter in the game is
varied, we find a transition where one player is able to get his
maximum theoretical score. By extending this model, we suggest that
the front velocity $v$ is selected by the nonlinear marginal
stability mechanism arising in some traveling wave problems for
which the profile decays exponentially, and for which standard
traveling wave theory applies.
\end{abstract}

\pacs{02.50.-r, 05.40.-a} \maketitle

Extreme value statistics of random variables has been long studied
by mathematicians \cite{Gumbel,Galambos} and physicists
\cite{KM,MK,DM}. In physics, it naturally arises when studying
thermodynamical properties of disordered systems \cite{DS}, and in
particular, the distribution of the ground state energy
\cite{MK,DM}.

If the considered random variables $E_1,E_2,...,E_N$ are
uncorrelated, the distribution of $E_{\rm min}=\min_i E_i$ or
$E_{\rm max}=\max_i E_i$ becomes universal for large $N$, once
properly scaled \cite{Gumbel,Galambos}. It takes the form of the
Gumbel, Fr\'echet or Weibull distribution depending on the asymptotic
properties of the distribution of the $E_i$'s. However, in the case
of strongly correlated random variables, there are no general
results, and it is usually a formidable task to access to the
distribution of $E_{\rm min}$ or $E_{\rm max}$.

In \cite{MK}, the authors study a simple model of a directed polymer
on a Cayley tree, inspired from the original work of \cite{DS}, but
focussing on the ground state properties (\textit{i.e.} zero
temperature). The simplest version of the model is defined on a
Cayley tree developed over $n$ generations, and with $Z$ branches
originating from each node. The polymer made of $n$ bonds starts
from the root node, and for a given path on the tree, the total
length of the polymer (assimilated to its energy) is
\begin{equation}
E_{\rm path}=\sum_{i\in{\rm path}} l_i.
\end{equation}
The elementary lengths $l_i$ are quenched random variables
associated to each bond of the tree and independently drawn from the
same random distribution $\rho(l)$. The hierarchical structure of
the Cayley tree induces strong correlations between the different
possible energies (or lengths) of the polymer. One is then
interested in the distribution of the minimal energy $E_{\rm min}$,
\emph{i.e.} the ground state energy. Because of these strong
correlations, the ground state energy distribution in not expected
to fall into one of the three universality classes arising in the
case of independent random variables \cite{MK,DM}, the best known of
them being the Gumbel distribution \cite{Gumbel,Galambos}. In the
special case of the binary distribution
\begin{equation}
\rho(l)=p\,\delta_{l,1}+(1-p)\,\delta_{l,0},\quad p\in[0,1],\label{score}
\end{equation}
the authors of \cite{MK} obtained an unbinding transition when
$p>p_c=1-Z^{-1}$, where the polymer goes from a finite length to an
extensive length $ \langle E_{\rm min}\rangle\sim v(p)n$. In
addition, the distribution of $E_{\rm min}$ has a traveling front
form
\begin{equation}
P(E_{\rm min},n)=f(E_{\rm min}-v(p)n),
\end{equation}
where $f(z)$ decays exponentially fast for large negative argument.
This last property and the general theory of traveling waves
\cite{TV1,TV2,TV3,TV4,TV5} lead to a simple selection mechanism for
the front velocity $v(p)$ (linear marginal stability; see
hereafter).

In the present work, we define a game theoretical model, directly
inspired by this directed polymer model. Although our model lacks
any thermodynamical reference, it is certainly related to other
optimization problems, where the notions of extreme value statistics
and traveling front arise \cite{KM}.

Two players $A$ and $B$ play an alternating game of duration $n$,
with player $A$ starting the game. When it is his turn to play, a
player has a choice of $Z$ possible moves. Hence, the map of all
possible game histories has the structure of a Cayley tree with $Z$
branches originating from any node, and of length $2n$. The $i$-th
move by player $A$ brings him the additional score $a_i$, whereas
the next play by player $B$ add the value $b_i$ to the score of
player $A$. The score of player $B$ is defined as the opposite of
that of player $A$. The elementary scores $a_i$ and $b_i$ are
quenched random variables independently drawn from the same
distribution $\rho$. Ultimately, the final score of player $A$ is
\begin{equation}
E_{\rm path}=\sum_{i\in{\rm path}} a_i+b_i.
\end{equation}
The goal of player $A$ is to maximize its final score, whereas
player $B$ will do his best to select his plays in order to minimize
the score of player $A$, and hence maximizing his own score. The two
players have an \textit{a priori} knowledge of the game tree
structure so that the final score of player $A$ is defined as
\begin{equation}
S_n=\max_{{\rm available~choices~of}\, A}\,\min_{{\rm
available~choices~of}\, B} E_{\rm path}.\label{mm1}
\end{equation}
From now, we specialize to the case $Z=2$, although our results can
be easily extended to any $Z$. Moreover, we restrain ourselves to
the elementary score distribution given by Eq.~(\ref{score}). It
should be emphasized that the players do not pick their next play in
order just to maximize its local outcome (\textit{i.e.} $A$ picking
its next move among available branches with $a_i=1$ or $B$ picking
the minimum available $b_i$). If the players were adopting such a
simple depth-0 strategy, which would be their natural approach if
they did not have the prior knowledge of the $a_i$'s and $b_i$'s
distribution over the tree, the final score of player $A$ would be
simply the sum of $n$ independent variables of mean $p^2+2p(1-p)$
($A$ picks a branch with $a_i=1$, if there is one available), and
$n$ variables of mean $p^2$ ($B$ picks a branch with $b_i=0$, if
there is one available). Then the distribution of $S_n$ would be a
Gaussian (of width $\sigma\sim \sqrt{n}$), and mean $\langle
S_n\rangle=v_0(p)n$, with
\begin{equation}
v_0(p)=p^2+2p(1-p)+p^2=2p.\label{v0}
\end{equation}
Note that this result is identical to the score velocity obtained if
the players had picked their move at random: the depth-0 strategies
of both players exactly annihilate. Instead, having a global view of
the game tree, the players will try to direct the game into
favorable branches for them, in order to maximize their final score,
even if they may have sometimes to pick an unfavorable local move
($a_i=0$ for player $A$, $b_i=1$ for player $B$) in order to achieve
their goal. For $p>1/2$, there are more bonds with $a_i=1$ or
$b_i=1$, so that we expect that the objective of player $A$ should
be easier to achieve than that of player $B$. Hence, we anticipate
that $\langle S_n\rangle=v(p)n$, with
\begin{equation}
v(p)\geq v_0(p)=2p,\quad p\geq \frac{1}{2}.\label{bound}
\end{equation}
In the opposite case $p<1/2$, the above inequality is obviously
reversed. In fact, by exchanging the roles of $A$ and $B$ (and
neglecting the fact that $A$ starts the game, for large $n$), it is
clear that one has the symmetry relation \cite{MK},
\begin{equation}
v(p)+v(1-p)=2.\label{sym}
\end{equation}
In addition, one has the trivial constraints,
\begin{equation}
v(0)=0,\quad v(1/2)=1,\quad v(1)=2,\label{cond}
\end{equation}
which are consistent with Eq.~(\ref{sym}).

An intermediate strategy to the ones presented above corresponds to
players having only a partial view  of the game tree up to a finite
depth. For instance, if the players have the knowledge of there next
available move, and of their opponent's ensuing options, they should
adopt the following depth-1 strategy:
\begin{itemize}
\item[$\bullet$] Player $A$: if the options of player $A$
are equal (both $a_i=0$ or 1), $A$ picks the branch for which the
number of $b_i$ equal to 1 (when it will be the turn of $B$ to
play) is maximal. If only one branch corresponds to $a_i=1$, $A$
choses this move.
\item[$\bullet$] Player $B$: if the options of player $B$
are equal (both $b_i=0$ or 1), $B$ picks the branch for which the
number of $a_{i+1}$ equal to 1 is minimal. If only one branch
corresponds to $b_i=0$, $B$ picks this move.
\end{itemize}
After a elementary but cumbersome calculation, we find that the
score velocity $v_1(p)$ corresponding to this depth-1 strategy is
given by,
\begin{equation}
v_1(p)=2{p^2}{\frac{7 - 6p + 4{p^2} - 14{p^3} + 14{p^4} -
        4{p^5} }{1 + 2p + 6{p^2} - 16{p^3} + 8{p^4}}}.
\end{equation}
One has $v_1(p)\geq v_0(p)$ for $p\geq 1/2$, and $v_1(p)$ satisfies
the symmetry relation of Eq.~(\ref{sym}) and the conditions of
Eq.~(\ref{cond}).
\begin{figure}[htbp]
\centerline{
\includegraphics[width=7cm,angle=0]{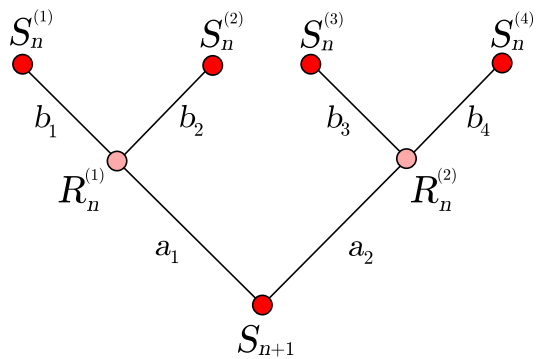}}
\caption[]{(Color online) $S_{n+1}$ can be obtained recursively from four
optimized scores $S_n^{(k)}$ ($k=1,2,3,4$), and finding the next
optimized move from player $B$ and then from player $A$.}
\label{tree}
\end{figure}
For higher but finite strategy depth, an analytical treatment
becomes extremely complicated.

Let us now move back to our model, where both players have a global
knowledge of the game tree (infinite depth). Obtaining the (not
necessarily unique) optimal path realizing both players antagonist
goals can be achieved by using the recursive \textit{minimax}
algorithm \cite{minim}, which gives a more precise meaning to
Eq.~(\ref{mm1}). Let us assume that we have generated four instances
of optimized scores $S_n^{(k)}$ ($k=1,2,3,4$) on four independent
games of length $n$ (with the initial condition $S_n^{(k)}=0$, for
$n=0$). In order to construct an optimized score for a game of
length $n+1$, we first generate two intermediate scores including
the previous move of player $B$ (see Fig.~\ref{tree}),
\begin{eqnarray}
R_n^{(1)}&=&\min\left(S_n^{(1)}+b_1,S_n^{(2)}+b_2\right),\nonumber\\
R_n^{(2)}&=&\min\left(S_n^{(3)}+b_3,S_n^{(4)}+b_4\right).\label{recb}
\end{eqnarray}
Then the final score is obtained by optimizing the first move of
player $A$ over his two possible plays, $a_1$ and $a_2$ (see
Fig.~\ref{tree}):
\begin{equation}
S_{n+1}=\max\left(R_n^{(1)}+a_1, R_n^{(2)}+a_2\right).\label{reca}
\end{equation}
Using Eqs.~(\ref{recb},\ref{reca}), we can derive the corresponding
recursion relations for the cumulative distribution of $S_n$ and
$R_n$,
\begin{eqnarray}
P_n(m)&=&{\rm Prob}\left(S_n\leq m\right),\nonumber\\
Q_n(m)&=&{\rm Prob}\left(R_n\leq m\right),\label{cumu}
\end{eqnarray}
and starting from the initial condition $P_n(m)=1$ for $m\geq 0$,
and $P_n(m)=0$ for $m<0$. Defining $q=1-p$, we find
\begin{eqnarray}
Q_n(m)&=& 1-\left(1-q P_n(m)-p P_n(m-1)\right)^2\label{recdist1}\\
P_{n+1}(m)&=& \left(q Q_n(m)+p Q_n(m-1)\right)^2.\label{recdist2}
\end{eqnarray}
The intermediate distribution $Q_n(m)$ can be eliminated by
inserting Eq.~(\ref{recdist1}) into Eq.~(\ref{recdist2}), leading to
a single recursion relation between $P_{n+1}$ and $P_n$. The
probability density of $S_n$ is defined as
\begin{equation}
p_n(m)=P_n(m)-P_n(m-1).
\end{equation}

We look for a traveling wave form for $P_n(m)$
\begin{equation}
P_n(m)=F(m-\langle S_n\rangle),\quad \langle S_n\rangle=v(p)n,
\end{equation}
with the boundary conditions $F(z)\to 1$, for $z\to+\infty$, and
$F(z)\to 0$, for $z\to-\infty$. The probability density of $S_n$ has
a similar traveling wave form, associated to the hull function
$f(z)$:
\begin{equation}
p_n(m)=f(m-\langle S_n\rangle),\quad f(z)=F(z)-F(z-1).\label{dists}
\end{equation}
Inserting this \emph{ansatz} into
Eqs.~(\ref{recdist1},\ref{recdist2}), we find that $F$ satisfies the
following functional equation
\begin{eqnarray}
\sqrt{F(z-v)}&=&1-q\left[1-q F(z)-  p F(z-1)\right]^2\nonumber\\
&-&p\left[1-q F(z-1)- p F(z-2)\right]^2,\label{funceq}
\end{eqnarray}
where we have used the shorthand notation $v$ for $v(p)$. By
retaining the leading contributions in Eq.~(\ref{funceq}) for
$z\to-\infty$, and for $v>0$, we find
\begin{equation}
F(z-v)\sim 4q^4 F^2(z),
\end{equation}
which leads to the double exponential asymptotics
\begin{equation}
F(z)\sim f(z)\sim \frac{1}{4q^4}\exp\left(-\alpha_-
2^{\frac{|z|}{v}}\right),\label{asymm}
\end{equation}
where $\alpha_->0$ is an unknown $p$-dependent constant. Similarly,
in the opposite limit $z\to+\infty$, and assuming $v<2$, the
functional equation Eq.~(\ref{funceq}) reduces to
\begin{equation}
1-F(z-v)\sim 2p^3 (1-F(z-2))^2,\label{Fsmall}
\end{equation}
which again leads to a double exponential decay
\begin{equation}
1-F(z-1)\sim f(z)\sim \frac{1}{2p^3}\exp\left(-\alpha_+
2^{\frac{z}{2-v}}\right),\label{asymp}
\end{equation}
where $\alpha_+>0$ is some $p$-dependent constant. Hence, and
contrary to the standard traveling wave theory \cite{KM,MK,DM},
where the traveling front exponential decay for $z\to-\infty$ or
$z\to+\infty$ permits the determination of the front velocity,
$v(p)$ remains so far undetermined. Here, the double exponential
decay obtained on both sides results from the \emph{minimax}
constraint, instead of the usual \emph{min} (or \emph{max})
constraint imposed when considering the ground state energy or the
minimum (or maximum) path length distribution \cite{MK,DM}. This
fast decay of $f(z)$ for $z\to\pm\infty$ and the traveling wave form
of Eq.~(\ref{dists}) ensure that $\left\langle
(S_n-v(p)n)^2\right\rangle$ remains bounded when $n\to+\infty$.

\begin{figure}[htbp]
\centerline{
\includegraphics[width=7cm,angle=0]{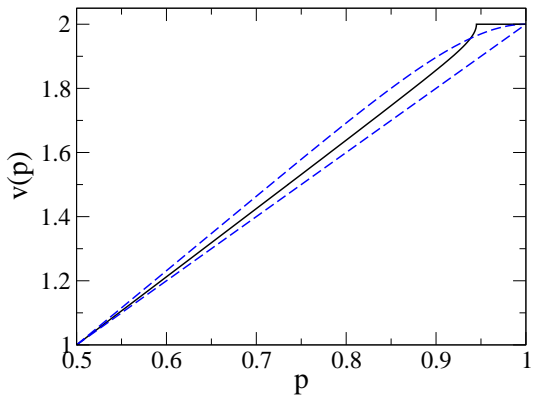}}
\caption[]{(Color online) We plot the score velocity front $v(p)$ (full line), the
lower bound $v_0(p)=2p$ (lower dashed line), and $v_1(p)$, obtained
when the players adopt a depth-1 strategy (upper dashed line). For $p<1/2$, $v(p)$ is
obtained by symmetry, using Eq.~(\ref{sym}). The heuristic
expression of Eq.~(\ref{fit}) cannot be distinguished from the
numerical data at this scale.} \label{Vp}
\end{figure}

$v(p)$ can still be determined numerically, from its definition
$\langle S_n\rangle=v(p)n$. The results are shown on Fig.~\ref{Vp},
along with $v_0(p)$ and $v_1(p)$ which correspond to depth-0 and
depth-1 strategies respectively. The main feature of $v(p)$ is the
existence of a critical value of $p$ (denoted $p_c$), above which
the score front velocity is $v(p)=2$ (note that one also has
$v'_1(1)=0$). Moreover, and as mentioned above, $v_0(p)$ is a lower
bound of $v(p)$. Finally, for $p$ close to $1/2$, $v(p)$ grows
linearly with $p$, with
\begin{eqnarray}
v'_0(1/2)&=&2,\nonumber\\ v'(1/2)&\approx &2.123(1),\nonumber\\
v'_1(1/2)&=&\frac{37}{16}=2.3125.
\end{eqnarray}
In the symmetric case $p=1/2$, we find that $\langle
S_n\rangle=n+\sigma_0$, where $\sigma_0= 0.14291695...$ is a
strictly positive constant, illustrating the slight advantage that
$A$ gains from starting the game.

Let us give a physical explanation for the occurrence of this
transition. As $p>1/2$ increases, the number of paths along which
all the $a_i$'s and $b_i$'s are equal to 1 grows exponentially.
Indeed, the probability of having such a path is $p^{2n}$, so that
their total number in the tree is of order $2^{2n}{\times} p^{2n}$. The
existence of this transition shows that for $p>p_c$, player $A$ is
able to chose his moves in order to force the outcome of the game to
follow one of these path, with probability unity, as $n\to\infty$.
Symmetrically, for $p<1-p_c$, player $B$ will find enough branches
along which most $a_i$'s and $b_i$'s are equal to 0, in order to
enforce that the front velocity remains zero in this regime,
consistently with the symmetry relation of Eq.~(\ref{sym}).

This transition can be understood analytically, by studying the
stability of a traveling front of velocity $v(p)=2$. Noting that
$P_n(2n)=Q_n(2n+1)=1$ (since $S_n\leq 2n$ and $R_n\leq 2n+1$), we
consider the dynamics of $u_n=P_n(2n-1)\leq 1$. From
Eqs.~(\ref{recdist1},\ref{recdist2}), $u_n$ satisfies the recursion
relation
\begin{equation}
u_{n+1}=\left[1-p^3(1-u_n)^2 \right]^2,\label{recu}
\end{equation}
with $u_0=0$. $u_n=P_n(2n-1)=1$ is an obvious fixed point of
Eq.~(\ref{recu}), corresponding to the case $v(p)<2$. If this fixed
point is selected, and after linearizing Eq.~(\ref{recu}), we find
that
\begin{equation}
\ln (1-P_n(2n-1))\sim  -2^{n},\label{asympu}
\end{equation}
which is fully consistent with Eq.~(\ref{asymp}), with
$z=2n-1-v(p)n$. However, the recursion relation of Eq.~(\ref{recu})
has three other fixed points $(x_-,x_+,x_0)$, which can be obtained
analytically by solving a third order polynomial equation, leading
to cumbersome expressions. A detailed analysis shows that $x_0$ is
always real, with $x_0>1$. This fixed point is unphysical and
necessarily unstable. The two other fixed points are real for $p\geq
p_c$ ($x_-<x_+$), with
\begin{equation}
p_c=\frac{3}{4}.\,2^{1/3}=0.94494...,\label{pc},
\end{equation}
and complex conjugates for $p<p_c$. Moreover, one finds that $x_-$
is maximal for $p=p_c$, at which $x_\pm=1/9$. Finally, a stability
analysis shows that $x_-$ is the only stable fixed point for
$p_c<p\leq 1$. Hence, we conclude that $P_n(2n-1)$ converges
(exponentially fast) to $x_-$ for $p_c<p\leq 1$, and that the
distribution of $S_n$ is peaked near $m=2n$ and decays as a double
exponential for $m\ll 2n$ (as given by Eq.~(\ref{asymm})), leading
to $v(p)=2$. The obtained value of $p_c$ is in perfect agreement
with the numerical results for $v(p)$ plotted on Fig.~\ref{Vp}.
Close to $p=1$, $x_-\sim 9(1-p)^2\to 0$, and up to second order in
$(1-p)$, the distribution of $S_n$ is thus given by
\begin{equation}
p_n(2n)=1-9(1-p)^2,\quad p_n(2n-1)=9(1-p)^2.
\end{equation}
Note that if both players adopt a finite depth strategy, this
transition does not occur, as illustrated in Fig.~\ref{Vp} in the
case of depth-0 and depth-1 strategies considered above. By adopting
a short-sighted strategy, player $A$ (respectively $B$) cannot
direct, with probability 1, the sequence of plays to a branch of the
tree with a density unity of playing options $a_i=b_i=1$
(respectively $a_i=b_i=0$).

Finally, for $p$ below but close to $p_c$, we obtain a very
convincing fit of $v(p)$ to the functional form
\begin{equation}
v(p)=2-c\,(p_c-p)^{1/2}+...,\label{behav}
\end{equation}
with $c\approx 0.50(1)$, leading to an infinite slope for $v(p)$ at
$p=p_c^-$, as found numerically on Fig.~\ref{Vp}. In fact, for $p\geq 1/2$,
we find that the simple heuristic functional form
\begin{equation}
v(p)=2-2\left(\frac{(p_c-p)(1-p)}{2p_c-1}\right)^{1/2},\label{fit}
\end{equation}
fits the data with a relative accuracy better that 0.1\,\%,
comparable although slightly higher than the estimated numerical
error bars of the data. This functional form ensures that $v(1/2)=1$
and that the behavior of Eq.~(\ref{behav}) is reproduced, and leads
to the heuristic values,
\begin{equation}
v'(0)=2.12099...,\quad c=0.49748...,
\end{equation}
in good agreement with the numerical estimates presented above.

\begin{figure}[htbp]
\centerline{
\includegraphics[width=7cm,angle=0]{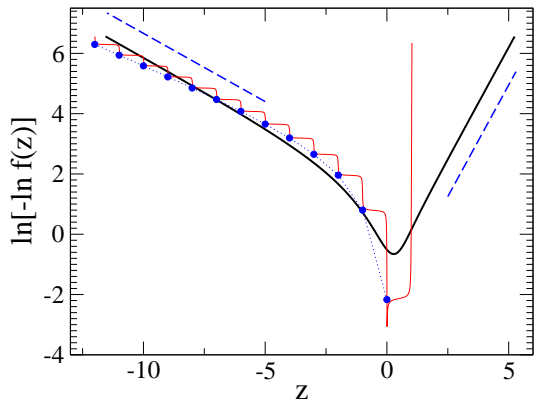}
} \caption[]{(Color online) We plot the hull function $f(z)$ for $p=0.94495>p_c$,
for which $v(p)=2$, and which is defined on negative integer values
of $z$ (dots linked by a dotted line). For $p=0.9449$ slightly below $p_c$
($v(p)=1.99684...$), the hull function is continuous but exhibits
smooth steps at integer values of $z$ (corresponding thin line).
Finally, for $p=0.75$ ($v(p)=1.53146...$), we plot the continuous
hull function (thick line) along with the predicted asymptotics of
Eqs.~(\ref{asymm},\ref{asymp}) (dashed lines).} \label{hull}
\end{figure}

Let us now address the properties of the hull function $f(z)$, and
its cumulative sum $F(z)$. First of all, if for a given $p$, the
corresponding $v(p)$ happens to be a rational number
$v(p)=\alpha/\beta$ ($\alpha$ and $\beta$ being mutually prime),
Eq.~(\ref{funceq}) implies that the hull function is only defined on
the discrete set of fractions of the form $k/\beta$. This is in
particular the case for $p=1/2$ ($v(1/2)=1$), $p>p_c$ ($v(p)=2$),
and $p<1-p_c$ ($v(p)=0$). On the other hand, when $v(p)$ is
irrational, the set of points of the form $z=m-v(p)n$ is dense on
the real axis, and $f(z)$ is a continuous function defined on the
real axis. As $v(p)$ approaches $v(p_c)=2$ from below, the hull
function $f(z)$ develops steps which blend into discontinuities as
$p\to p_c$. This property is illustrated on Fig.~\ref{hull}, along
with the asymptotics obtained in Eqs.~(\ref{asymm},\ref{asymp}).

We now extend our original model in order to gain some insight about
the velocity selection mechanism. This is achieved by modifying the
model so that the standard theory of front propagation will apply.
In this $(A,\varepsilon)$-model, player $A$ always follows its best
strategy, while player $B$ follows the depth-0 strategy (picking a
branch with $b_i=0$ if available) with probability $\varepsilon$ and
its optimal strategy with probability $1-\varepsilon$. The original
model is hence recovered for $\varepsilon=0$. The recursion relation
of Eq.~(\ref{recdist1}) now becomes
\begin{eqnarray}
Q_n(m)=&& (1-\varepsilon)
\left(1-\left(1-q P_n(m)-p P_n(m-1)\right)^2\right)\nonumber\\
&&+\varepsilon\left(\left(q^2+2pq \right)P_n(m)+p^2 P_n(m-1)\right),
\label{recdist3}
\end{eqnarray}
while Eq.~(\ref{recdist2}) remains unchanged. We denote the
associated front velocity by $v_A(p,\varepsilon)$. Note that if $B$
were playing purely randomly, the terms $q^2+2pq$ and $p^2$ in
Eq.~(\ref{recdist3}) must be replaced respectively by $q$ and $p$.
This model has exactly the same qualitative properties as the
$(A,\varepsilon)$-model, on which we hence concentrate, since player
$B$ adopts a more intelligent strategy (some results obtained for
the random model will be mentioned in passing). In the
$(B,\varepsilon)$-model, one exchanges the role of players $A$ and
$B$, and Eq.~(\ref{recdist2}) is changed into
\begin{eqnarray}
P_{n+1}(m)= (1-\varepsilon)\left(q Q_n(m)+p Q_n(m-1)\right)^2\nonumber\\
+\varepsilon\left(q^2 Q_n(m)+\left(p^2+2pq \right) Q_n(m-1)\right),\label{recdist4}
\end{eqnarray}
whereas Eq.~(\ref{recdist1}) still holds. When exchanging the role
of the two players, the associated front velocity
$v_B(p,\varepsilon)$ satisfies the symmetry relation,
\begin{equation}
v_A(p,\varepsilon)+v_B(1-p,\varepsilon)=2,\label{symA}
\end{equation}
which reduces to Eq.~(\ref{sym}), when $\varepsilon=0$. However, for
$\varepsilon>0$, $v_A(p,\varepsilon)$ and $v_B(1-p,\varepsilon)$ do
not obey the symmetry relation of Eq.~(\ref{sym}). In addition, we
have the exact inequalities,
\begin{equation}
v_B(p,\varepsilon)\leq v(p)\leq v_A(p,\varepsilon),\label{eqAB}
\end{equation}
since players $B$ and $A$ are respectively playing less optimally in
the $(A,\varepsilon)$ and $(B,\varepsilon)$-models than in the
original model.

Because of the symmetry relation of Eq.~(\ref{symA}), we focus on
the $(A,\varepsilon)$-model, and denote the associated velocity
simply by $v$. The $(A,\varepsilon)$-model main interest lies in the
fact that for $\varepsilon>0$, $\bar P_n(m)=1-P_n(m)$ decays
exponentially for $m\gg vn$, so that the standard mechanisms of
front velocity selection do apply (see below). When $\bar P_n(m)\ll
1$, the recursions of Eqs.~(\ref{recdist2},\ref{recdist3}) indeed
lead to
\begin{eqnarray}
\bar P_{n+1}(m)=2\varepsilon\left[(1+p)(1-p)^2\bar P_n(m)\right.\nonumber\\
\left.+p(1-p) (2 p+1)\bar P_n(m-1)+p^3\bar P_n(m-2)\right],\label{smallP}
\end{eqnarray}
or equivalently, the front profile $\bar F(z)=1- F(z)$ satisfies
\begin{eqnarray}
\bar F(z-v)=2\varepsilon\left[(1+p)(1-p)^2\bar F(z)\right.\nonumber\\
\left.+p(1-p) (2 p+1)\bar F(z-1)+p^3\bar F(z-v)\right],\label{smallF}
\end{eqnarray}
to be compared to Eq.~(\ref{Fsmall}), for the original model. The
simple \emph{ansatz} $\bar F(z)\sim\exp(-\lambda z)$ leads to the
dispersion relation
\begin{equation}
v(\lambda)=\frac{1}{\lambda}\ln\left[2\varepsilon
\left(1-p+p{\rm e}^\lambda\right)
\left(1-p^2+p^2{\rm e}^\lambda\right)\right]\label{vlambda},
\end{equation}
where the decay rate $\lambda$ is so far undetermined.

Let us now summarize the main known mechanisms of velocity front
selection \cite{TV1,TV2,TV3,TV4,TV5} for exponentially fast decaying
profiles. In many physical cases, including those studied in \cite{MK,DM}, a
linear marginal stability (LMS) argument shows that the selected
front velocity corresponds to the minimum velocity $v_{\rm min}$
allowed by the dispersion relation $v(\lambda)$, associated to the
decay rate $\lambda_{\rm min}$. However, in some other cases
\cite{TV3,TV4,TV5}, a bigger velocity is selected by a nonlinear
marginal stability (NLMS) mechanism. Without entering into too much
details, let us briefly explicit this point. Consider the
large $z$ asymptotics of a solution of
the full nonlinear problem associated to the velocity $v$,
\begin{equation}
\bar F(z)\sim A_1(v)\,{\rm e}^{-\lambda_1(v)z}+
A_2(v)\,{\rm e}^{-\lambda_2(v)z}+...\label{expex},
\end{equation}
with $\lambda_1(v)<\lambda_{\rm min}$ given by the dispersion
relation derived from linear analysis. Note that the above linear
analysis does not grant access to $A_1(v)$, not to mention the
correction term proportional to $A_2(v)$. Now, if there exists a
velocity $v_*>v_{\rm min}$ for which $A_1(v_*)=0$, $\bar F(z)$ will
decay more sharply with rate $\lambda_2(v_*)$, which is necessarily
another root of the dispersion relation, with
$\lambda_2(v_*)>\lambda_{\rm min}$. It can  be shown that all
traveling fronts with velocity less than $v_*$ are then unstable
against invasion by a profile of velocity $v_*$, which leads to the
selection of the velocity front $v_*$, instead of $v_{\rm min}$
\cite{TV3,TV4,TV5}. In practice, there are very few examples for
which the transition between a linear and a nonlinear marginal
stability scenario can be analytically identified, since it requires
in general a full solution of the profile associated to a velocity
$v$, in order to obtain $A_1(v)$. To the knowledge of the author,
all such tractable examples concern traveling front in  the spatial
and temporal continuum \cite{TV3,TV4,TV5}, like for instance,
\begin{equation}
\frac{\partial P}{\partial t}=\frac{\partial^2 P}{\partial x^2}+P(1-P)(1+\kappa P),
\end{equation}
for $\kappa\geq 1$. In this case \cite{TV3,TV5},
$v(\lambda)=\lambda+\lambda^{-1}$, so that $v_{\rm min}=2$ and
$\lambda_{\rm min}=1$. $v_{\rm min}$ is selected for
$1\leq\kappa\leq 2$, whereas
$v=v_*=(\kappa/2)^{1/2}+(\kappa/2)^{-1/2}$ (with
$\lambda_1=(\kappa/2)^{-1/2}$ and $\lambda_2=(\kappa/2)^{1/2}$), for
$\kappa>2$.

Returning to our $(A,\varepsilon)$-model, we find that a non trivial
$v_{\rm min}$ exists for any $\varepsilon>1/2$. It is obtained by
first finding $\lambda_{\rm min}$, the unique real positive solution
of
\begin{equation}
v'(\lambda_{\rm min})=0,\label{l0min}
\end{equation}
and setting $v_{\rm min}=v(\lambda_{\rm min})$ in
Eq.~(\ref{vlambda}). In particular, we find that $v_{\rm min}=2$,
for $p>p_c$, with
\begin{equation}
p_c=(2\varepsilon)^{-1/3}.\label{pclinear}
\end{equation}
In the case $\varepsilon=1$, when player $B$ always adopts the
depth-0 strategy, we find $p_c=2^{-1/3}=0.7937005...$ Hence, we
obtain the same kind of transition as in the original model, where
player $A$ is able to get its maximum theoretical score. However,
since $B$ has a short-sighted strategy, we do not observe a
transition to $v=0$ for small but non zero $p$, as obtained in the
original model for $p=1-p_c$. We actually find $v_{\rm
min}\sim-\frac{\ln (2\varepsilon)}{\ln (p)}$, when $p\to 0$.  Note
that if $B$ plays randomly instead of adopting the depth-0 strategy,
we obtain the dispersion relation
\begin{equation}
v(\lambda)=\frac{1}{\lambda}\ln\left[2\varepsilon
\left(1-p+p{\rm e}^\lambda\right)
^2\right]\label{vlambda00},
\end{equation}
and $p_c=(2\varepsilon)^{-1/2}$.

However, for a given $p$, we find numerically that the velocity
given by the LMS mechanism $v_{\rm min}$ is only selected for
$\varepsilon\geq\varepsilon_c(p)$, so that the results of
Eqs.~(\ref{vlambda},\ref{l0min},\ref{pclinear}) are only valid for
$\varepsilon$ close enough to 1. For
$1/2<\varepsilon<\varepsilon_c(p)<1$, and although a non trivial
$v_{\rm min}$ does exist, we find $v>v_{\rm min}$. This strongly
suggests the relevance of the NLMS mechanism in this case.
Unfortunately, for $1/2<\varepsilon<\varepsilon_c(p)$ and a given
$v$, there is very little hope to obtain an analytical solution of
the corresponding full nonlinear equation for $F(z)$, in order to
apply the NLMS criterion explained above. Likewise, for
$\varepsilon<1/2$, the minimal positive velocity is $v_{\rm min}=0$
($v(\lambda=0)=-\infty$), and the prospect of an analytical solution
appears even bleaker. Note however that $v$ and the associated decay
rate $\lambda$ are still related by the dispersion relation of
Eq.~(\ref{vlambda}).

\begin{figure}[htbp]
\centerline{
\includegraphics[width=7cm,angle=0]{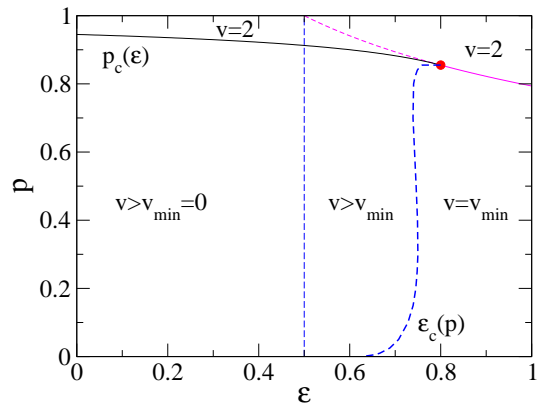}
} \caption[]{(Color online) We plot the exact $p_c(\varepsilon)$
above which $v_A(p,\varepsilon)=2$ (full line), whose LMS and NLMS
expressions are respectively given by Eq.~(\ref{pclinear}) and
Eq.~(\ref{pcnl}). We also plot the numerical estimate of
$\varepsilon_c(p)$ (thick dashed line), the boundary between the LMS
and NLMS domains of application. These two curves cross at
$(\varepsilon_c,p_c)=\left(\frac{4}{5}, \frac{5^{1/3}}{2}\right)$
(dot). We also identify three domains according to the relation
between the observed front velocity $v$ and the LMS velocity $v_{\rm
min}$.} \label{peps}
\end{figure}
On the bright side, the full line $p_c(\varepsilon)$ can be
determined exactly, by studying the dynamics of $u_n=1-P_n(2n-1)$,
in the same spirit as in the case $\varepsilon=0$. We find that
$u_n$ satisfies the exact recursion relation
\begin{eqnarray}
u_{n+1}&\equiv& g_{p,\varepsilon}(u_n)=z_n(2-z_n),\\
z_n&=&p^3 u_n(\varepsilon+(1-\varepsilon)u_n),\label{une}
\end{eqnarray}
with $u_0=1$. We then determine the value of $p_c(\varepsilon)$
above which there exists a stable non trivial fixed point
$u_*(\varepsilon)\ne 0$. For $\varepsilon\geq 4/5$, we find that
$p_c(\varepsilon)$ is indeed given by the LMS argument, leading to
the result of Eq.~(\ref{pclinear}). On the other hand, for
$0\leq\varepsilon<4/5$, a regime were the NLMS mechanism should be
relevant, $p_c(\varepsilon)$ and $u_*(\varepsilon)$ are determined
in the same spirit as in the case $\varepsilon=0$, by imposing that
$g_{p_c,\varepsilon}(u_*)/u_*-1=g'_{p_c,\varepsilon}(u_*)-1=0$. We
find
\begin{eqnarray}
p_c(\varepsilon)=\left[\frac{(1-\varepsilon)^{1/2}(4-\varepsilon)^{3/2}-
8+7\varepsilon+\varepsilon^2}{2\varepsilon^2}\right]^{1/3},\label{pcnl}
\end{eqnarray}
which goes smoothly to the result of Eq.~(\ref{pc}), when
$\varepsilon\to 0$.

Interestingly, this analysis provides the exact value of
$\varepsilon_c$ for the corresponding value of
$p=p_c(\varepsilon_c)=(2\varepsilon_c)^{-1/3}$, at the transition
between the LMS and NLMS regimes. We thus find
\begin{equation}
\varepsilon_c\left(p=\frac{5^{1/3}}{2}=0.854988...\right)=\frac{4}{5}.\label{epsc}
\end{equation}
If $B$ plays randomly instead of adopting the depth-0 strategy, one
obtains
\begin{equation}
\varepsilon_c\left(p=\frac{1+\sqrt{5}}{4}=0.809017...\right)=
3-\sqrt{5}=0.763932...\label{epsc0}
\end{equation}

\begin{figure}[htbp]
\centerline{
\includegraphics[width=7cm,angle=0]{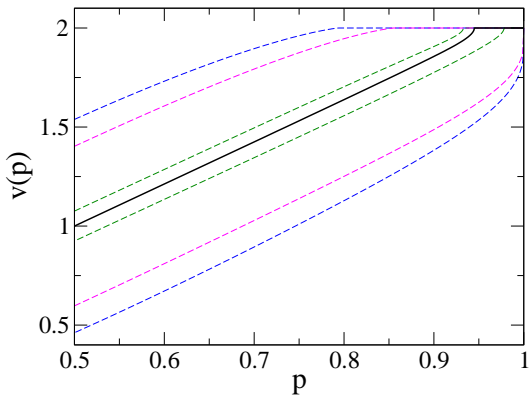}
} \caption[]{(Color online) We plot the velocities
$v_A(p,\varepsilon)$ (three upper dashed lines;
$\varepsilon=1,\,4/5,\,1/4$ from top) and  $v_B(p,\varepsilon)$
(three lower dashed lines; $\varepsilon=1,\,4/5,\,1/4$ from
bottom). The full line corresponds to $v(p)$ for the original model
($\varepsilon=0$).} \label{vpeps}
\end{figure}

In Fig~\ref{peps}, we plot our exact result for $p_c(\varepsilon)$
and a numerical estimate of $\varepsilon_c(p)$, the boundary between
the LMS and NLMS domains of application. In Fig~\ref{vpeps}, we plot
$v_A(p,\varepsilon)$ and $v_B(p,\varepsilon)$ for $\varepsilon=1$,
where $B$ and $A$ are respectively adopting the depth-0 strategy,
for $\varepsilon=4/5$, the smallest $\varepsilon$ for which the LMS
mechanism applies for all $p$, and for $\varepsilon=1/4$, for which
the NLMS mechanism holds for all $p$. We observe numerically that
$v_A(p,\varepsilon)$ and $v_B(p,\varepsilon)$ converge smoothly to
$v(p)$ as $\varepsilon\to 0$.

Let us finally address the subleading corrections to the average
score $\langle S_n\rangle$. The LMS mechanism implies \cite{TV1,TV4}
that
\begin{equation}
\langle S_n\rangle=v_{\rm min}n-\frac{3}{2\lambda_{\rm min}}\ln n+...
\end{equation}
In the $(\varepsilon,p)$ regime where the LMS mechanism applies, we
have confirmed numerically the occurrence of this logarithmic
correction, as well as its magnitude. In the NLMS regime, we find
instead that the next correction to $\langle S_n\rangle=vn$ is
constant. Quite generally, this result can be justified analytically
whenever $v>v_{\rm min}$, in particular when the NLMS mechanism
applies \cite{CS}. This property was exploited in order to obtain
the numerical estimate of $\varepsilon_c(p)$ shown in
Fig~\ref{peps}, and this criterion is found to be fully consistent
with defining $\varepsilon_c(p)$ as the value of $\varepsilon$ for
which $v$ becomes equal to the velocity $v_{\rm min}$ selected by
the LMS mechanism (see
Eqs.~(\ref{vlambda},\ref{l0min},\ref{pclinear})).

In the present work, we have defined a simple two-player game
inspired by a model of directed polymer on the Cayley tree. The fact
that the two players have antagonist goals is reminiscent of the
notion of \emph{frustration} quite common in disordered physical
systems. In our model, this frustration originates from the
\emph{minimax} constraint, which is, however, quite uncommon in
physics. As a consequence, the present model has no thermodynamical
interpretation. We found that the score distribution develops a
traveling wave form, with the hull function unusually decaying
superexponentially for large negative and positive arguments. We
have justified analytically the occurrence of a transition, across
which a player can obtain his maximum theoretical score, whatever
the strategy of the other player. Contrary to systems for which the
standard traveling wave theory applies, we did not succeed in
understanding analytically the process which leads to the selection
of the velocity front. However, after studying an extension of the
original model, we strongly suggest that the selection mechanism is
related to nonlinear marginal stability, arising in some traveling
wave problems for which the profile decays exponentially.

\smallskip
\acknowledgments I am very grateful to Satya Majumdar for fruitful
discussions. I also wish to thank one referee for suggesting
studying a model where only one player is adopting the optimal
strategy whereas the other plays randomly.

\end{document}